\documentclass[12pt]{article}

\usepackage[backend=biber,sorting=none,style=ieee,citestyle=numeric-comp,isbn=false,dashed=false,url=false]{biblatex}
\usepackage{graphicx}
\usepackage{xcolor}
\usepackage{amsmath}
\usepackage{amssymb}
\usepackage{mdframed}
\usepackage[breaklinks=true]{hyperref}
\usepackage{authblk}
\usepackage[utf8]{inputenc}
\usepackage[T1]{fontenc} 
\usepackage{url}

\graphicspath{{figures/}}

\definecolor{arxivc}{RGB}{22,72,145}
\def\arxivtourl#1{http://arxiv.org/abs/#1}      
\DeclareFieldFormat{eprint:arxiv}{%
    \ifhyperref
        {\color{arxivc}\href{\arxivtourl#1}{[\nolinkurl{arxiv:#1}]}}
        {\nolinkurl{#1}}}

\graphicspath{{figures/}}

\newcommand{\mo}{\mathcal{O}}
\newcommand{\code}[1]{\texttt{#1}}

\title{\LARGE Simulating CDT quantum gravity}

\author[1]{J. Brunekreef\thanks{jorenb@gmail.com}}
\author[2,3]{A. G\"orlich\thanks{andrzej.goerlich@uj.edu.pl}}
\author[1]{R. Loll\thanks{r.loll@science.ru.nl}}

\affil[1]{Institute for Mathematics, Astrophysics and Particle Physics, Radboud University, Heyendaalseweg 135, 6525 AJ Nijmegen, The Netherlands \vspace{3pt}}
\affil[2]{Jagiellonian University, Institute of Theoretical Physics, \mbox{{\L}ojasiewicza 11, Krak\'ow, PL 30-348, Poland}\vspace{3pt}}
\affil[3]{Jagiellonian University, Mark Kac Center for Complex Systems Research,  \mbox{{\L}ojasiewicza 11, Krak\'ow, PL 30-348, Poland}}

\begin{document}

\maketitle






\vspace{0.8cm}

\begin{center}
{\bf Abstract}
\end{center}
We provide a hands-on introduction to Monte Carlo simulations in nonperturbative lattice quantum gravity, formulated in terms of Causal Dynamical Triangulations (CDT). We describe explicitly the implementation of Monte Carlo moves and the associated detailed-balance equations in two and three spacetime dimensions. We discuss how to optimize data storage and retrieval, which are nontrivial due to the dynamical nature of the lattices, and how to reconstruct the full geometry from selected stored data. Various aspects of the simulation, including tuning, thermalization and the measurement of observables are also treated. An associated open-source C++ implementation code is freely available online.

\vspace{12pt}
\noindent

\newpage

\tableofcontents

\newpage
\section{Introduction and motivation}
In our search for a full-fledged theory of quantum gravity beyond perturbation theory, the importance of nonperturbative
computational and numerical tools can hardly be overemphasized. While elsewhere in physics the use of such tools is commonplace when
analyzing the behaviour of strongly coupled systems, from condensed matter to nonperturbative QCD to strong gravity, this is much less the case in 
quantum gravity, although it is a prime candidate for the application of computational methods. This is now changing, with
a growing body of evidence and accordingly appreciation in the field that numerical techniques can deliver valuable quantitative insights into
the physics of quantum gravity on Planckian scales, which currently cannot be obtained by other means. However, there is also a lack of familiarity 
with nonperturbative tools like lattice or Monte Carlo methods as such, and what it takes in practical terms to apply them to gravity.
Our contribution aims to give some hands-on guidance for how to set up such tools in systems of quantum gravity,
to help lower the threshold for new practitioners and strengthen this indispensable pillar of quantum gravity research further.

\subsection{Putting quantum gravity on the lattice}
As a consequence of the particular field content and symmetry structure of gravity, which captures the intrinsic dynamics of spacetime, 
it has been a highly challenging task to develop quantitative, quantum field-theoretic methods that can describe its quantum 
theory in a nonperturbative manner.\footnote{Note that our starting point are the degrees of freedom of general relativity, 
and we do not invoke extra ingredients and symmetries, or concepts beyond those of covariant quantum field theory.}  
Although the beginnings of the field in the late 1970s were very much inspired by the early successes of lattice QCD (see \cite{loll1998discrete} for a review 
and evaluation of
such quantum gravity approaches), significant difficulties stand in the way of a straightforward adaptation  
of the techniques of lattice gauge field theory to gravity. Among them are the need for a Wick rotation for general curved geometries, the unboundedness of
the gravitational action and the requirements of diffeomorphism invariance and background independence. 

To cut a long story short, the good news is that in the intervening time enormous progress has been made on these issues, and ways have been found
to address and control them. A very important ingredient is the use of so-called random geometry,
a way to represent the nonperturbative gravitational path integral as the
continuum limit of a well-defined sum over an ensemble of piecewise flat geometries. This methodology, also called dynamical triangulation(s), 
emerged in the 1980s in the context of noncritical string theory, to describe the dynamics of curved two-dimensional world sheets of Euclidean metric signature \cite{david1985planar,ambjorn1997quantumb,budd2023lessons}. A key point is that the fixed lattice structure used in standard 
lattice regularizations of quantum and statistical field theories is dropped in favour of \textit{dynamical} lattices, obtained by assembling a set of
identical geometric building blocks in all possible ways into metric spaces with curvature. As we will see later, the dynamical nature of these lattices 
has nontrivial implications for how data are stored and manipulated in numerical simulations. 

\subsection{CDT as a computational tool}
In pursuit of a fundamental, physical theory of quantum gravity in 4D, this methodology was developed further into Causal Dynamical 
Triangulations (CDT), which incorporates the Lorentzian nature of the curved spacetimes contributing to the path integral. 
The associated Lorentzian path integral was solved analytically in 2D and lies in a distinct
universality class from its Euclidean counterpart \cite{ambjorn1998nonperturbative}. Crucial from a calculational perspective is the fact that in two 
\textit{and} higher dimensions, the gravitational path integral based on CDT is amenable to powerful Markov chain Monte Carlo techniques. 
In particular, it gives us a rare window on the nonperturbative dynamics of quantum gravity in four spacetime dimensions, 
a physical realm about which essentially nothing was known, from any perspective!  

The extensive exploration of this nonperturbative sector with the help of Monte Carlo simulations of the 4D CDT path integral has so far been highly
successful in producing new, quantitative results on quantum observables describing the Planckian behaviour of quantum spacetime, 
including evidence 
that they are compatible with a well-defined classical limit. We will not cover these results and their physical ramifications here, which have been 
discussed in numerous articles, and refer the interested reader to several recent overview and review articles on CDT quantum gravity  
\cite{ambjorn2012nonperturbative,loll2019quantum,ambjorn2021cdt,ambjorn2023lattice,loll2023quantum}. 

For our present purposes we will treat CDT simply as a computational tool for investigating nonperturbative quantum gravity, 
analogous to
how lattice gauge theory is used to investigate the nonperturbative properties of QCD. This is justified by the simplicity and robustness
of the formulation and the nature of the results obtained so far. CDT provides a
straightforward regularization of the local geometric degrees of freedom of general relativity, which are then summed over in a regularized version of
the gravitational Feynman path integral, without any exotic ingredients or assumptions. Moreover, in the physically relevant 4D implementation,
quantum observables evaluated on spacetimes (``universes'') with a diameter of about 20$\,\ell_{\rm Pl}$ exhibit properties that are
universal and physically meaningful, in the sense of being compatible with semiclassical outcomes or expectations from other approaches 
(like in the case of the spectral dimension observable \cite{ambjorn2005spectral,rechenberger2012phasediagram,carlip2017dimension}). 

All of this suggests that CDT provides a valid \textit{effective} description of 4D quantum gravity near the Planck scale,
which is a very nontrivial statement. Note that this is independent of the question of whether CDT also provides an ultraviolet completion of quantum gravity, or
whether such a completion can only be realized at transplanckian length scales in terms of some other set of (yet more) fundamental excitations. 
A possible realization of the former scenario would be a confirmation in the CDT formulation of the presence of an ultraviolet fixed point of the 
renormalization group \cite{ambjorn2014renormalization,ambjorn2020renormalization,ambjorn2023lattice}. 
The phase diagram of CDT indeed contains several lines of second-order phase transitions \cite{ambjorn2011secondorder,ambjorn2017new}, which are promising candidates for such fixed points. 
The investigation of renormalization group flows and the search for fixed points are the subject of ongoing research, but not directly relevant
to our exposition below. 

\subsection{Some theory basics}
\label{sec:basis}
As indicated earlier, we will focus on the computational and implementational aspects of the CDT simulations, and keep the inclusion of other background
material to a minimum.  
We start by recalling some key quantities from the continuum theory that will be studied in the lattice discretization. 
The formal gravitational path integral as a function of the Lorentzian metric tensor $g_{\mu\nu}(x)$ on a manifold $M$ is given by
\begin{equation}
Z[G_{\rm N},\Lambda]=\int {\cal D}[g] \, {\rm e}^{iS[g]},\;\;\;\; S[g] =\frac{1}{16\pi G_{\rm N}}\int_M\! d^4x\sqrt{|\det g|} \, (R[g]-2\Lambda),
\label{picont}
\end{equation}
where the functional integration is over diffeomorphism equivalence classes $[g_{\mu\nu}]$ of metrics and $S[g]$ is the Einstein-Hilbert action,
which depends on Newton's constant $G_{\rm N}$ and the cosmological constant $\Lambda$.
For definiteness, we have stated the path integral in 4D; in the examples below we will work with the corresponding simplified models
in two and three spacetime dimensions. CDT gives us a precise regularization prescription that converts the formal expression (\ref{picont}) into a finite
and mathematically well-defined quantity, whose continuum limit and renormalization can be studied systematically with the help of numerical
techniques. This involves a specification of the regularized configuration space in terms of Lorentzian triangulations, the path integral measure
and the action, as well as performing the analytical continuation (``Wick rotation'') that is part of the CDT formulation 
(see \cite{ambjorn2001dynamically,ambjorn2012nonperturbative} for details). 
The latter step is needed to
convert the complex path integral into a real partition function amenable to Monte Carlo simulations. In a formal continuum language, this
partition function has the form
\begin{equation}
Z[G_{\rm N},\Lambda]=\int {\cal D}[g_E] \, {\rm e}^{-S_E[g_E]},
\label{pieucl}
\end{equation}
where $S_E$ is the Euclidean Einstein-Hilbert action depending on the Wick-rotated geometry $g_E$, and by a slight abuse of
notation we continue to use the symbol $Z$ for the Euclidean version of the path integral. The expectation value of a diffeomorphism-invariant
observable $\cal O$ is then computed according to
\begin{equation}
\langle{\cal O}\rangle =\frac{1}{Z} \int {\cal D}[g_E] \, {\cal O}[g_E] \,  {\rm e}^{-S_E[g_E]}.
\label{exval}
\end{equation}
The lattice counterpart of the path integral (\ref{pieucl}) in the CDT formulation is given by a sum over (Wick-rotated) causal triangulations $T$,
\begin{equation}
Z[G_{\rm N},\Lambda]=\sum_{\rm causal\, T}\frac{1}{C(T)}\, {\rm e}^{-S_E[T]},
\label{picdt}
\end{equation}
where the integer $C(T)$ is the size of the automorphism group of the triangulation $T$, and $G_{\rm N}$ and $\Lambda$ should now be interpreted as
bare coupling constants, as usual in lattice field theory. The standard implementation of the Einstein-Hilbert action used in (\ref{picdt}) is the so-called Regge action,
which expresses the volume and curvature integrals in terms of the edge lengths of the triangulation $T$. The explicit expressions in 2D and 3D
will be given in the relevant sections below. The formula for the expectation value of a geometric operator $\cal O$ in CDT is
\begin{equation}
\langle{\cal O}\rangle =\frac{1}{Z} \sum_{\rm causal\, T}\frac{1}{C(T)}\, {\cal O}[T]\, {\rm e}^{-S_E[T]},
\label{exvalcdt}
\end{equation}
with the normalization factor $1/Z$ given in terms of the CDT path integral $Z$ of eq.\ (\ref{picdt}). 

In spacetime dimension two, the infinite sum in (\ref{picdt})
can be performed and renormalized explicitly, yielding a continuum theory of two-dimensional Lorentzian quantum geometry \cite{ambjorn1998nonperturbative}.
Also the spectral properties of some selected observables can be computed exactly.  
However, computing the expectation values of observables in any dimension analytically is in general intractable. 
In these cases, we can use computer simulations to sample ensembles of triangulations with a finite number of building blocks, 
allowing us to compute estimators of such expectation values. 

Details of the theoretical underpinnings of the regularized path integral of CDT, including the construction of the simplicial action, 
the Wick rotation and the transfer matrix, are described in \cite{ambjorn2001dynamically,ambjorn2012nonperturbative}.
Note that we will work throughout with configuration spaces of simplicial manifolds. A \emph{simplicial mani\-fold} of dimension $d$ is characterized by
the property that the link of each $m$-simplex $\sigma^m$ has the topology of the sphere $S^{d-m-1}$. 
The \emph{link} of $\sigma^m$ consists of all faces (sub-simplices) $f$ of all simplices
in the star of $\sigma^m$ which have a trivial intersection with $\sigma^m$, i.e.\ $f \cap \sigma^m=\emptyset$, 
and the \emph{star} of $\sigma^m$ is the union of all simplices containing $\sigma^m$ \cite{ambjorn1997geometry}.\footnote{A ``link'' in the
technical sense defined here should not be confused with the generic name for a one-simplex, which we interchangeably refer to as a link or an edge.}
All our simulations contain checks to make sure that the manifold property is maintained during Monte Carlo moves. 
One could in principle relax these regularity conditions, by allowing for local degeneracies of some kind. However, beyond spacetime dimension
two little is known about how this affects the continuum limit of the models (see \cite{brunekreef2022phase} for an investigation in 3D CDT). 
For such generalized ensembles, the discussion of detailed balance and symmetry factors associated with relabellings will in general be subtle, 
and one must also pay attention to the size of the entropic contribution of the degenerate configurations in the path integral. 

In what follows, we will provide a hands-on introduction of the basic principles involved in CDT computer simulations, and explain in detail how these principles can be put into practice. In Sec.\ \ref{mc-sec:mcmrg} we discuss the general set-up of Markov chain Monte Carlo (MCMC) methods for random geometry and
describe the basic ingredients of CDT simulations in two and three dimensions. In Sec.\ \ref{sec:impl} we outline some of the difficulties one encounters when requiring the simulations to run efficiently, and how a particular method of storing simplices in memory can be used to address these issues. We also discuss the three stages of a simulation run: the tuning of the bare coupling constants to their pseudocritical values, the evolution of the initial geometry to a well-thermalized state, and the measurement of observables in the quantum geometry. This section is based heavily on our own implementation of CDT simulations in two and three dimensions. The associated C++ implementation code \cite{brunekreef2021jorenb,brunekreef2022jorenb} has been open-sourced and is freely available for use by quantum gravity enthousiasts. We end in Sec.\ \ref{sec:summ} with a short summary and outlook.

\section{Monte Carlo methods for random geometry}
\label{mc-sec:mcmrg}
Our aim is to generate a sample $X$ of an ensemble $\mathcal{T}$ of triangulations, with the probability distribution 
\begin{equation}
    P(T) = \frac{1}{Z}\,  e^{-S_E(T)}
    \label{mc-eq:prob-sample}
\end{equation}
for drawing a triangulation $T$ from the ensemble, where the partition function $Z$ was defined in (\ref{picdt}).
The sample mean of an observable $\mo$ over $X$ gives an estimator $\widehat{\langle \mo \rangle}$ of the ensemble average $\langle \mo \rangle$. In order to generate the sample $X$ we can make use of Markov chain Monte Carlo (MCMC) methods. The idea behind this approach is to construct a random walk (in the form of a Markov chain) in the space of triangulations, where we perform local updates to a geometry at each step of the walk. By enforcing certain conditions on this random walk, we can ensure that the probability of encountering a specific triangulation $T$ in the random walk approaches the probability distribution \eqref{mc-eq:prob-sample}. 

If we sample from this random walk with a sufficiently large number of local updates in between subsequent draws, we obtain a set of independent triangulations with the desired probability distribution. This set forms the sample $X$. Note that in the computer simulations we always work with \emph{labelled} triangulations, where triangulations related by a relabelling of building blocks correspond to the same physical geometry.
This requires us to introduce a combinatorial factor to correct for this overcounting, so that for a labelled triangulation $T$ we have
\begin{equation}
    P_l(T) = \frac{1}{Z} \frac{1}{N_0(T)!}\, e^{-S_E(T)},
    \label{mc-eq:prob-labeled}
\end{equation}
where $N_0(T)$ counts the number of vertices in $T$.\footnote{In what follows, we will use the notation ``$T$'' for  both labelled and unlabelled
triangu\-lations, where the correct interpretation should usually be clear from the context. In the explicit implementation of the Monte Carlo moves, we will highlight 
wherever the labelled character of a configuration is important.} We make the choice of labelling the vertices rather than any of the higher-dimensional 
simplex types. The latter can be described in terms of lists of vertex labels.

In practice, it is convenient to start the random walk from a simple triangulation constructed by hand. To avoid that this leads to a bias in the measurement results, one should minimize the dependence on this special choice of initial geometry. We implement this by letting 
the simulation run for a long time, such that the system can \emph{thermalize} (see Sec.\ \ref{ssec:simul} for further details). When this thermalization phase is completed, we have obtained an independent triangulation that can be used as the initial configuration for a series of measurements. In between subsequent measurements, we then perform a large number of local updates, which we group together in batches, so-called \emph{sweeps}. The size of a sweep should be taken large enough to ensure that significant autocorrelations with the previously measured geometry have disappeared. Since larger sweeps require longer simulation times, one typically has to strike a balance between reducing autocorrelation influences and increasing execution time. Later in this section we discuss how to determine appropriate lengths for the thermalization phases and sweeps.
We refer the interested reader to \cite{newman1999monte} for a more detailed explanation of the MCMC approach and associated practical issues like autocorrelation effects. Next, we will discuss briefly the main elements relevant for our purposes.

\subsection{Random walks and detailed balance}
The first condition we should enforce on the random walk is \emph{ergodicity}, meaning that all triangulations in the geometric ensemble $\mathcal{T}$ can be reached in a finite number of steps. Without ergodicity, we would be restricted to a subset of the ensemble, potentially making the estimators of quantum expectation values unreliable. The second condition relates to the transition probabilities of a single step in the random walk. Let us denote the probability for transitioning from a triangulation $T$ to a new triangulation $T'$ in a single step by $p(T \to T')$. For the probability distribution of the random walk to converge to \eqref{mc-eq:prob-sample}, it is sufficient to demand that the transition probabilities satisfy the condition of \emph{detailed balance},
\begin{equation}
    P_l(T) p(T \to T') = P_l(T') p(T' \to T).
    \label{mc-eq:detailed-balance}
\end{equation}
The transition probabilities are nonzero only between triangulations that are related by the basic local moves which we consider for the ensemble.
For the allowed transitions a solution to the detailed balance equation \eqref{mc-eq:detailed-balance} is
\begin{align}
    \frac{P_l(T)}{P_l(T')} = \frac{p(T' \to T)}{p(T \to T')}.
    \label{mc-eq:transition-probs}
\end{align}
We furthermore allow for trivial transitions $T \to T$ from a triangulation to itself, i.e.\ $p(T \to T) \neq 0$. Labelling every step of the walk with an integer external ``time'' parameter $n$, the random walk consists of a sequence of triangulations $T_n$. The updating procedure then works as follows. At step $n$ of the random walk, we have that $T_n \! =\! T$. We select one of the allowed local updates $T \to T'$ with a \emph{selection probability} $g(T \to T')$. This proposed update, also called a \emph{move}, is subsequently accepted with probability $A(T \to T')$, and rejected otherwise. The probability $A(T \to T')$ for a move to be accepted is called the \emph{acceptance ratio}. In case it is accepted, we set the next triangulation in the sequence to $T_{n+1}\! =\! T'$. If the move is rejected, the triangulation is unchanged and we set $T_{n+1}\! =\! T$. 

The transition probability for the move $T \to T'$ can be decomposed as
\begin{equation}
    p(T \to T') = g(T \to T') A(T \to T'). 
\label{transprob}    
\end{equation}
The selection probability $g(T \to T')$ depends on the set of basic moves, their implementation, and the details of the triangulation $T$ at hand. Subsequently, the acceptance ratio $A(T \to T')$ should be chosen in such a way to ensure detailed balance. A simple choice is the one made in the \emph{Metropolis-Hastings algorithm} \cite{metropolis1953equation,hastings1970monte}, and amounts to setting
\begin{align}
    A(T \to T') &= \textrm{min}\left(1, \frac{g(T' \to T) P_l(T')}{g(T \to T') P_l(T)} \right), \label{mc-eq:acceptance-ratios_1} \\
    A(T' \to T) &= \textrm{min}\left(1, \frac{g(T \to T') P_l(T)}{g(T' \to T) P_l(T')} \right).
    \label{mc-eq:acceptance-ratios_2}
\end{align}
It is instructive to see how this works in the two example scenarios of two- and three-dimensional CDT, and how the choice of the move implementation affects the selection probabilities and associated acceptance ratios. In what follows, we will denote by an \emph{$(n,m)$-move} a geometric update 
of a $d$-dimensional triangulation, for which a local neighbourhood consisting of $n$ $d$-simplices before the move is transformed to a neighbourhood of  
$m$ $d$-simplices after the move.

\subsection{Metropolis algorithm for 2D CDT}
\label{mc-sec:metropolis-2d}
We start by computing the probability $P_l(T)$ of encountering a given labelled triangulation $T$ in the ensemble of 2D CDT with toroidal topology. 
Since the curvature term of the Einstein-Hilbert action in two dimensions is a topological invariant, we can drop it from the path integral, leaving only 
the cosmological-constant term. Eq.\ \eqref{mc-eq:prob-labeled} becomes 
\begin{equation}
    P_l(T) = \frac{1}{Z} \frac{1}{N_0(T)!}\, e^{-\lambda N_2(T)},
    \label{labweight}
\end{equation}
where $\lambda$ is the bare cosmological constant and $N_2(T)$ is the discrete volume (number of two-simplices) of the triangulation $T$.
Note that for toroidal topology the number of triangles is always even, and we have $N_2(T)\! =\! 2 N_0 (T)$.

\begin{figure}[t]
    \centerline{\scalebox{0.42}{\rotatebox{0}{\includegraphics{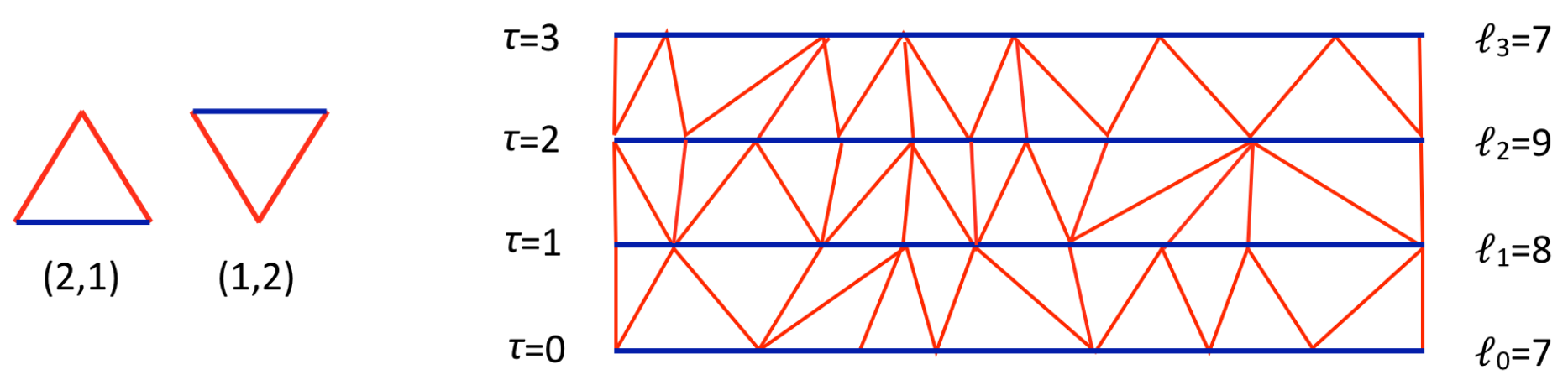}}}}
    \caption{Left: fundamental building block of 2D CDT with two different time orientations, the (2,1)-simplex and the (1,2)-simplex. Spacelike edges are blue,
    and timelike ones red. Right: part of a CDT configuration, given in terms of a sequence of spatial universes of length $\ell_\tau$ at integer time steps $\tau$, with (2,1)- and (1,2)-triangles interpolating
    between slices; vertical boundaries on the left and right should be identified.}
    \label{CDTstrip}
\end{figure}

Recall that CDT configurations are assembled from $N_2$ identical Minkows\-kian triangles with two timelike edges of equal length and
one spacelike edge.\footnote{As is customary, we refer to time- and spacelike edges despite the fact that we have already performed
the Wick rotation present in CDT and all edges have spacelike lengths.} This fundamental building block appears with two distinct time orientations, 
which we call the (2,1)- and the (1,2)-simplex (see Fig.\ \ref{CDTstrip}, left).\footnote{This conforms with standard notation, which we will use throughout,
where a $(k,l)$-simplex is a simplex that shares $k$ of its vertices with a spatial slice at some integer time $\tau$, and the remaining $l$ vertices with
the spatial slice at time $\tau\! +\! 1$.} A given simplicial geometry $T$ consists of a sequence of one-dimensional spatial slices of variable integer length $\ell$ 
(the number of its spacelike
edges), each one representing a compact spatial universe of $S^1$-topology at discrete ``proper time'' $\tau\! =\! 0,1,2,3,\dots$ (see Fig.\ \ref{CDTstrip}, right).
Strip configurations, consisting of sequences of up- and down-triangles, interpolate between adjacent spatial slices, as in the example shown. 
Expression (\ref{labweight}) will be used to compute the left-hand side of eq.\ \eqref{mc-eq:transition-probs} for a given move and its inverse. 

\begin{figure}[t]
    \centering
    \includegraphics[width=0.4\textwidth]{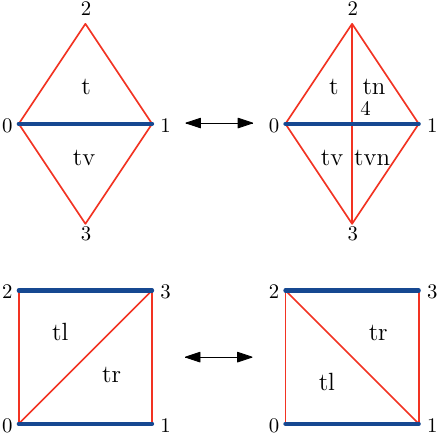}
    \caption{The two basic sets of moves used in 2D CDT: \code{add}/\code{delete} move (top) and \code{flip} move (bottom). The labels of the vertices are for identification purposes. }
    \label{mc-fig:2d-cdt-moves}
\end{figure}

Two simple local moves (and their inverses), illustrated in Fig.\ \ref{mc-fig:2d-cdt-moves}, suffice to construct an ergodic random walk in the space 
of two-dimensional CDT geometries with a fixed number of spatial slices.
The first move, the so-called $(2,4)$-move or \code{add} move, takes two adjacent triangles that share a spacelike edge, and replaces them by four triangles,
as shown in the figure.
Its inverse, the $(4,2)$-move or \code{delete} move, takes four triangles that share a single vertex of order four\footnote{In dimension 2, 
the order of a vertex $v$ is defined as the number of triangles that contain $v$.}, and replaces them by two triangles. 
The second move, the so-called $(2,2)$-move or \code{flip} move, takes two adjacent triangles with opposite time orientations that share a timelike edge, 
and flips this edge to the opposite diagonal, as shown in Fig.\ \ref{mc-fig:2d-cdt-moves}. This move is its own inverse. In the following subsections, 
we will discuss the detailed implementation of either set of moves.

\subsubsection{The \code{add} and \code{delete} moves}
\label{sub:adddelete}
We start by considering the (2,4)-move or \code{add} move and its inverse, the (4,2)-move or \code{delete} move. 
Performing the \code{add} move on a labelled triangulation 
$T$ with $N_2$ triangles results in a labelled triangulation $T'$ with $N_2 + 2$ triangles, by splitting a spacelike link in two and adding a new vertex and two timelike links in the middle. 
A simple method for implementing it consists of  an operation that adds two triangles to $T$, followed by a relabelling of vertices. 
We start by randomly selecting a triangle \code{t} in $T$, with uniform probability $1/N_2$ per triangle,
and finding the unique triangle \code{tv} that shares a spacelike link with \code{t} and therefore is its \code{v}ertical neighbour. 
The two triangles \code{t}, \code{tv} can also be described by their constituent vertices, as indicated in Fig.\ \ref{mc-fig:2d-cdt-moves}. 
Using the convention that vertices are listed in counterclockwise order, we have \code{t} $\! =\! (012)$ and \code{tv} $\! =\! (031)$. 
Note that the two-simplex \code{t} in our example is of type (2,1); in case \code{t} is of type (1,2), the move should proceed analogously. 

The first part of the proposed move consists of inserting a new vertex, labelled 4 in the figure, adding two new triangles \code{tn} 
$\! =\! (412)$ and \code{tvn} $\! =\! (431)$ to the right of the triangles \code{t} and \code{tv}, 
and reassigning the vertices of the latter to \code{t} $\! =\! (042)$ and \code{tv} $\! =\! (034)$. 
The resulting labelled triangulation $\tilde{T}'$ contains $N_2 + 2$ triangles and $N_2/2 + 1$ vertices.

However, as mentioned above, the complete move $T\rightarrow T'$ proceeds in two steps.
In general, assuming that the vertices of $T$ were labelled 1 to $N_0$ (these labels are generally different from the ones shown in the figure),
the new vertex \code{v} in $\tilde{T}'$ will be assigned the label $N_0 + 1$. As a second step, in order to not distinguish this vertex in terms of its labelling 
(as the highest label), 
we randomly select an arbitrary vertex $\code{v}'$ of $\tilde{T}'$ and swap its label with that of the new vertex \code{v}, to obtain the final, labelled triangulation
we call $T'$. There are $N_0+1$ ways of selecting a vertex $\code{v}'$ in $\tilde{T}'$; 
if the selected vertex happens to be \code{v} itself, the label swap is trivial. 
Furthermore, since an initial selection of the triangle \code{tv} instead of \code{t} would have led to the same labelled triangulation, 
we must multiply the probability of going from $T$ to $T'$ by a factor of two. Taking into account that $N_2\! =\! 2 N_0$, the overall \emph{selection probability} 
for this move is therefore given by
\begin{equation}
g(T \to T') = \frac{2}{N_2} \cdot \frac{1}{N_0 + 1} =  \frac{1}{N_0 \cdot (N_0+1)},
\label{moveprob}
\end{equation}
where $N_0$ is understood as $N_0(T)$.
We do not need to perform any checks before proposing this move, since it can always be carried out for any input triangle \code{t} in the geometry. 
This is not the case for its inverse, the (4,2)-move, since it requires that the vertex shared by the four initial triangles be of order four. We now present two distinct ways of implementing the \code{delete} move, which will influence the acceptance ratios. Since we used an initial labelled triangulation $T$ 
on which we performed the \code{add} move, we take the resulting labelled triangulation $T'$ as a starting point for the \code{delete} move in order to 
compute the relevant terms in the acceptance ratios \eqref{mc-eq:acceptance-ratios_1} and \eqref{mc-eq:acceptance-ratios_2}.

\paragraph{Method I: blind guessing.}
The simplest approach amounts to randomly selecting a vertex \code{v} in $T'$ with uniform probability $\tfrac{1}{N_0(T')} = \tfrac{1}{N_0(T)+1}$. If \code{v} is not of order four, we do not perform any update on the geometry, and let the next triangulation in the random walk again be $T'$. 
On the other hand, if \code{v} is of order four we do either of two things: if the label of \code{v} is \emph{not} $N_0+1$, we swap it with the label of the 
vertex $\code{v}'$ in $T'$ which does carry the label $N_0+1$. Otherwise, if the label of the selected vertex \code{v} happens to be $N_0+1$, 
no relabelling is performed.
We then proceed to remove the vertex \code{v} alongside its label $N_0+1$ from the triangulation, as well as the pair of triangles to the right of \code{v} 
that contain \code{v}. In terms of Fig.\ \ref{mc-fig:2d-cdt-moves}, where the vertex \code{v} carries the label 4, we 
identify the four triangles that contain \code{v}, delete the triangles $(412), (431)$, and reassign the vertices of the other two triangles
according to $(042) \to (012), (034) \to (031)$. 
This combined move only brings us back to the labelled triangulation $T$ we started from if the vertex \code{v} we selected in $T'$ is the one that 
was added to the system by the \code{add} move discussed above. We conclude that the selection probability for the move $T' \to T$ 
is given by
\begin{equation}
g(T' \to T) = \frac{1}{N_0(T)+1}.
\label{invprob}
\end{equation} 
Substituting (\ref{moveprob}) and (\ref{invprob}) into eqs.\ \eqref{mc-eq:acceptance-ratios_1}, \eqref{mc-eq:acceptance-ratios_2} 
we find the acceptance ratios
\begin{align}
    A(T \to T') &= \textrm{min}\left(1, \frac{N_0(T)}{N_0(T)+1}\, e^{-2\lambda} \right), \label{accadd0} \\
    A(T' \to T) &= \textrm{min}\left(1, \frac{N_0(T)+1}{N_0(T)}\, e^{2 \lambda} \right). \label{accdel0}
\end{align}
The exponential terms are due to the change in the action, and would appear regardless of the choice of selection process. 
On the other hand, the ratio $N_0/(N_0+1)$ arises on combinatorial grounds, and is tied to the specific implementation of the moves.

\paragraph{Method II: bookkeeping.}
While the previous method is straightforward to implement, it will often result in no change to the triangulation, since the move is aborted 
whenever the selected vertex is not of order four. In a thermalized configuration, approximately $1/4$ of the vertices are of order four \cite{ambjorn1999new}, 
which means that on average the configuration is updated at most a quarter of the time a \code{delete} move is proposed. 
We may expect that we can do better by keeping track of the order-four vertices throughout the simulation, and selecting from this set 
as input for the \code{delete} move. 
The move is then always possible, and it remains to compute the corresponding selection probabilities and acceptance ratios. 
The selection probability for the \code{add} move is unchanged, but the selection probability for the \code{delete} move now becomes
\begin{equation} 
g(T' \to T) = \frac{1}{N_\textit{vf}\,(T) + 1},
\label{invprobalt}
\end{equation}
where $N_\textit{vf}\,(T)$ is the number of vertices of order four in the original triangulation $T$. For the corresponding acceptance ratios, we find
\begin{align}
    A(T \to T') &= \textrm{min}\left(1, \frac{N_0(T)}{N_\textit{vf}\,(T)+1}\, e^{-2\lambda} \right),\label{accadd} \\
    A(T' \to T) &= \textrm{min}\left(1, \frac{N_\textit{vf}\,(T)+1}{N_0(T)}\, e^{2 \lambda} \right).\label{accdel}
\end{align}
In two-dimensional CDT, the critical value of the cosmological constant is $\lambda^{\textrm{crit}}\! =\! \ln 2$, and we typically perform simulations at this fixed value for $\lambda$. The acceptance ratio \eqref{accdel} for the \code{delete} move is therefore of order 1 for thermalized configurations, just as it was in \eqref{accdel0} for the ``blind guessing'' approach due to the minimization. On the other hand, the acceptance ratio \eqref{accadd} for the \code{add} move is now also of order 1, as opposed to approximately $1/4$ in \eqref{accadd0} for blind guessing. This increase in accepted \code{add} moves by a factor four is counterbalanced by the fact that the bookkeeping method always proposes a vertex suitable for deletion, whereas the blind guessing approach only did so for roughly $1/4$ of the cases (for thermalized triangulations). As a result, the transition probabilities \eqref{transprob} for the \code{add}/\code{delete} move pair are four times higher when using the bookkeeping approach. Furthermore, it is likely that this method provides an even larger efficiency boost in unthermalized systems, where the proportion of order-four vertices can be much lower than in the thermalized setting.



We should point out that the above arguments for improved transition rates hinged upon the fact that the fraction of order-four vertices is approximately constant in this model at the (unique) critical cosmological coupling. Similar relations are not necessarily present in higher-dimensional models, and it may be the case that bookkeeping-type approaches are only more efficient for certain ranges of the coupling parameters. 

We end the discussion on bookkeeping versus blind guessing approaches by pointing out that attaining high acceptance rates for the moves does not guarantee optimal execution times in practice. The reason is that a simple check like the one occurring in method I is computationally cheap, and it may be the case that the overhead incurred by bookkeeping-type methods slows down the simulation to an extent that offsets the potential benefits. In many situations, it may therefore \emph{a priori} be unclear which of the approaches provides the best performance, and one will need to carry out profiling runs in order to determine the optimal strategy for the purpose at hand.

\subsubsection{The \code{flip} move}
\label{sub:flip}
We will now discuss the implementation of the $(2,2)$-move or \code{flip} move, illustrated in Fig.\ \ref{mc-fig:2d-cdt-moves}. 
We could again consider both blind-guessing and bookkeeping approaches, but will restrict our attention to the latter, since this is the approach 
used in the codebase \cite{brunekreef2021jorenb} discussed in Sec.\ \ref{sec:impl} below. 
The \code{flip} move is its own inverse, which means that the inverse does not have to be considered separately.

Instead of keeping track of all vertices of order four, we now maintain a list of all triangles that have a neighbour of the opposite time orientation to 
their \emph{right}, i.e.\ a (2,1)-simplex paired with a (1,2)-simplex or vice versa. The timelike link shared by such a triangle pair can be flipped without destroying the sliced structure of the spacetime geometry. Since the number of triangles stays constant during a \code{flip} move, the probability $P_l(T)$ of 
eq.\ \eqref{mc-eq:prob-labeled} is also unchanged, and the left-hand side of eq.\ \eqref{mc-eq:transition-probs} is equal to 1. 

For a given triangulation $T$, we denote the number of triangles which have a triangle of the opposite orientation to their right by $N_\textit{tf}(T)$. 
The move consists in selecting one such triangle \code{tl} with uniform probability $1/N_\textit{tf}$, together with its right neighbour \code{tr}, and flipping
their shared timelike link to the opposite diagonal. 
For illustration, let us consider the labelled configuration shown in Fig.\ \ref{mc-fig:2d-cdt-moves}, bottom. 
We have \code{tl} $=(032)$ and \code{tr} $=(013)$, where \code{tl} is a (1,2)-simplex and \code{tr} a (2,1)-simplex.
We then propose to flip the shared timelike edge $(03)$, resulting in a new triangulation $T'$ where $(03)$ is replaced by the new timelike edge $(12)$. 
Starting from a given labelled triangulation $T$, the selection probability for this move is 
\begin{equation}
g(T \to T') = \frac{1}{N_\textit{tf}(T)}. 
\label{selflip}
\end{equation}
We should pay attention to the fact that the flip move can change the number $N_\textit{tf}$, depending on the neighbouring triangles of \code{tl} 
and \code{tr}. If the triangle \code{tll} to the left of \code{tl} was initially of the same type as \code{tl}, 
it will be of the opposite type after the move, which implies that $N_\textit{tf}$ increases by 1. Similarly, if \code{tr} and the right neighbour \code{trr} of \code{tr}
were of the same type initially, they will not be after the move, again increasing $N_\textit{tf}$ by 1. 
Conversely, $N_\textit{tf}$ decreases by 1 whenever \code{tl} and \code{tll} or \code{tr} and \code{trr} were initially of opposite types.
For a given labelled triangulation $T'$, the selection probability for moving back to $T$ with a subsequent \code{flip} move is then
given by $g(T' \to T) = 1/N_\textit{tf}(T')$.
As a result, we find that the acceptance ratio for the \code{flip} is
\begin{align}
    A(T \to T') &= \textrm{min}\left(1, \frac{N_\textit{tf}(T)} {N_\textit{tf}(T')} \right).
\end{align}

\subsection{Metropolis algorithm for 3D CDT}
\label{mc-sec:3d-cdt-moves}
The discussion for CDT in three spacetime dimensions proceeds along similar lines but, not surprisingly, the geometry of the local moves
is more complex. Our choice of topology is $S^1\times S^2$, with compact spherical spatial slices and a compactified time direction.
The probability to encounter a given labelled triangulation $T$ in the 3D CDT ensemble is
\begin{equation}
    P_l(T) = \frac{1}{Z} \frac{1}{N_0(T)!}\, e^{k_0 N_0(T) - k_3 N_3(T)},
\end{equation}
where the argument of the exponential function is given by (minus) the simplicial version of the three-dimensional Einstein-Hilbert action.
The coupling $k_0$ is proportional to the inverse bare Newton constant and $k_3$ depends linearly on the bare cosmological constant \cite{ambjorn2001nonperturbative}. 
There are now three types of moves (and their inverses), which we will describe in more detail below when computing the corresponding acceptance ratios. 

\begin{figure}[t]
    \centering
    \includegraphics[width=0.6\textwidth]{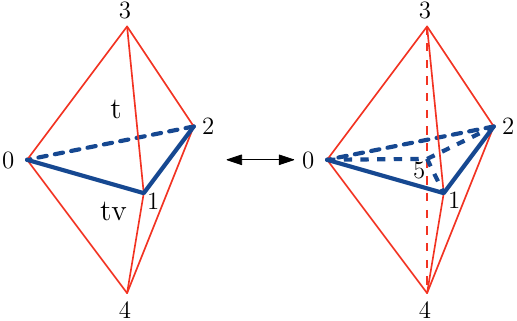}
    \caption{The \code{add} and \code{delete} moves in 3D CDT. }
    \label{mc-fig:3d-cdt-add-del}
\end{figure}

\begin{figure}[t]
    \centering
    \includegraphics[width=0.6\textwidth]{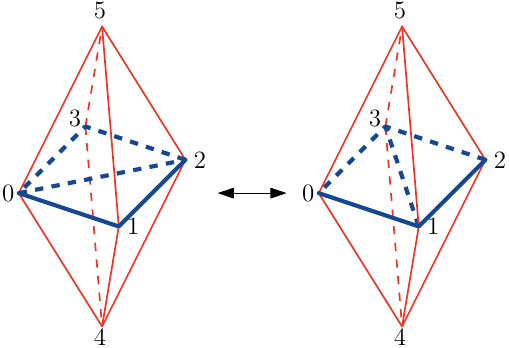}
    \caption{The \code{flip} move in 3D CDT.}
    \label{mc-fig:3d-cdt-flip}
\end{figure}

In brief, the first one is the $(2,6)$-move together with its inverse, the $(6,2)$-move, also referred to as \code{add} and \code{delete} moves 
(see Fig.\ \ref{mc-fig:3d-cdt-add-del}), since they are analogous to their 2D counterparts. The second one is the $(4,4)$-move or \code{flip} move, 
which is its own inverse. It flips a spatial edge as shown in Fig.\ \ref{mc-fig:3d-cdt-flip}. 
Lastly, there is the $(2,3)$-move together with its inverse, the $(3,2)$-move, also referred to as the \code{shift} and \code{inverse shift} (or \code{ishift}) moves.
The $(2,3)$-move is performed on an adjacent pair of a (3,1)- (or a (1,3)-)simplex and a (2,2)-simplex, and results in the overall addition of a
(2,2)-simplex, as illustrated in Fig.\ \ref{mc-fig:3d-cdt-shift-ishift}. The $(3,2)$-move reverses this process. 

When developing the codebase \cite{brunekreef2022jorenb} for simulating 3D CDT, we decided to implement all moves using 
blind-guessing approaches. It is in principle possible to include bookkeeping methods, by maintaining lists of special simplices where a particular move can be performed. However, the checks needed to keep these lists updated are nontrivial, and one must make sure that the lists are complete, to satisfy the detailed-balance condition. 
Since it is difficult to implement such checks properly in a computationally efficient way, we opted for blind-guessing approaches for the time being. 
Nevertheless, appropriate bookkeeping methods may well deliver substantial improvements to the execution time in some regions of the phase space. 
A closer exploration of this strategy will be important in the further development of the codebase.

\subsubsection{The \code{add} and \code{delete} moves}
Given a labelled triangulation $T$, 
we implement the (2,6)-move or \code{add} move by randomly selecting a (3,1)-simplex \code{t} with uniform probability $1/N_{31}(T)$,
where $N_{31}(T)$ denotes the number of (3,1)-simplices in $T$.
It shares a spatial triangle with a vertically adjacent (1,3)-simplex we will call \code{tv}. In terms of the labels in Fig.\ \ref{mc-fig:3d-cdt-add-del}, we
have $\code{t}\! =\! (0123)$, $\code{tv\! }=\! (0124)$, and the shared triangle is (012).
As part of the move $T\rightarrow T'$, a vertex $\code{v}$ is added at the centre of the shared triangle, which effectively adds two (3,1)- and 
two (1,3)-simplices to the system. 
The order of this vertex is six, which means there are six tetrahedra containing \code{v}. We also note that this move adds two 
spatial triangles to the spatial slice that contains \code{v}. With regard to the label assigned to \code{v}, we proceed exactly as in
subsection \ref{sub:adddelete} for the \code{add} move in two dimensions. We first assign to \code{v} the label $N_0+1$, and then swap it with the label of
an arbitrary vertex in the enlarged triangulation.
The resulting labelled triangulation $T'$ contains $N_0(T)+1$ vertices and $N_3(T) + 4$ tetrahedra, which implies that the ratio of the probabilities of
$T'$ and $T$ is given by
\begin{equation}
    \frac{P_l(T')}{P_l(T)} = \frac{1}{N_0(T)+1}\, e^{k_0 - 4k_3}.
\end{equation}
Following a similar reasoning as in two dimensions, the selection probability for the \code{add} move is then
\begin{equation}
g(T \to T') = \frac{1}{N_{31} \cdot (N_0 + 1)}.
\label{selprob3d}
\end{equation}
For the (6,2)-move or \code{delete} move we pick a random vertex \code{v} with uniform probability $1/(N_0(T)+1)$ from the labelled triangulation $T'$. 
Note that a vertex of order six is necessarily associated with a local configuration exactly as shown in Fig.\ \ref{mc-fig:3d-cdt-add-del}. Any such vertex 
is therefore a valid candidate for removal by the \code{delete} move. 
Like in the two-dimensional implementation, we do not update the geometry if the selected vertex is not of order six. 
Otherwise, if \code{v} is of order six, the proposed move consists of a label swap of \code{v} with the vertex that has the label $N_0+1$
(or no label swap if \code{v} already happens to have the label $N_0+1$), followed by a deletion of the vertex \code{v} 
and the removal of four of the six surrounding tetrahedra as shown in Fig.\ \ref{mc-fig:3d-cdt-add-del}. 
The probability that this brings us back to the triangulation $T$ is therefore $g(T' \to T) = 1/(N_0(T) + 1)$, since it only occurs when 
the selected vertex is the one that was added when performing the previously discussed \code{add} move.

Using the above results, we can compute the corresponding acceptance ratios for the \code{add} and \code{delete} moves, which are
\begin{align}
    A(T \to T') &= \textrm{min}\left(1, \frac{N_{31}(T)}{N_0(T)+1}\, e^{k_0 - 4 k_3} \right), \\
    A(T' \to T) &= \textrm{min}\left(1, \frac{N_0(T)+1}{N_{31}(T)}\, e^{-k_0 + 4 k_3} \right).
\end{align}
Taking into account that $N_0(t)\! =\! N_2(t)\! +\!\chi$ on a given spatial slice at time $\tau$, where $\chi$ denotes the Euler characteristic of the slice,
we have $N_{31}/(N_0\! +\! 1) \approx 2$ for large 3D triangulations $T$ with bounded time extension. 
A major difference with the two-dimensional case is that the couplings $k_0$ and $k_3$ are not fixed to unique (critical) values. 
The cosmological coupling must be tuned to its (pseudo-)critical value $k_3^{\textrm{crit}}(k_0)$ to obtain a continuum limit, but this value depends on $k_0$.
Furthermore, as was found in computer simulations of the model \cite{ambjorn2001nonperturbative,ambjorn2001computer,ambjorn20023d},
the gravitational coupling $k_0$ does not require further fine-tuning, but can be chosen from a wide range of values.
The acceptance ratios are affected by the choice of $k_0$, resulting in a different distribution on the triangulations in the ensemble.

\subsubsection{The \code{flip} move}
As input for the proposed (4,4)-move or \code{flip} move $T\!\rightarrow\! T'$, we pick a (3,1)-simplex \code{t} from the labelled triangulation $T$ with 
uniform probability $1/N_{31}(T)$. 
We subsequently select with uniform probability $1/3$ one of the three neighbouring tetrahedra that share a timelike face with \code{t}, and
denote it by \code{tn}. If \code{tn} is a (2,2)-simplex we abort the move. Conversely, if \code{tn} is a (3,1)-simplex, we flip the timelike triangle
shared by \code{t} and \code{tn}, which requires a similar rearrangement of the two (1,3)-simplices that share spatial triangles with 
\code{t} and \code{tn}. In Fig.\ \ref{mc-fig:3d-cdt-flip}, we have $\code{t}=(0125)$ and $\code{tn}=(0235)$, and the shared triangle (025)
is flipped to a new shared triangle (135), leading to the new three-simplices (0135) and (1235). At the same time, the (1,3)-simplices
(0124) and (0234) are replaced by the new (1,3)-simplices (0134) and (1234).
This induces a link flip to the opposite diagonal in the spatial ``square'' formed by the two spatial base triangles 
(012) and (023), as illustrated by the figure. No vertex relabellings are performed in constructing the final triangulation $T'$. 
Note also that there is an alternative way of getting from a given $T$ to $T'$, namely by first picking \code{tn} and then \code{t}.

There is no change in the number of (sub-)simplices during this move, which implies that $P_l(T)\! =\! P_l(T')$. 
Furthermore, the selection probabilities for a \code{flip} move and its inverse are the same, since we use identical blind guessing when
selecting the three-simplices where the flip is performed. As a consequence, the acceptance ratios are $A(T \to T')\! =\! A(T' \to T)\! =\! 1$, 
and the move is accepted whenever the conditions mentioned above are satisfied, and rejected otherwise. In line with statements we made earlier, it goes without
saying that a move is also rejected when it leads to a violation of the  simplicial manifold conditions. As a concrete example in 3D CDT,
take the flip move depicted in Fig.\ \ref{mc-fig:3d-cdt-flip} and consider the situation where another (3,1)-simplex is present, which has as its base
a spatial triangle (013), and has three timelike faces, the triangles (135), (015) and (035), where the latter two are shared with timelike triangles of
the (3,1)-simplices shown in the figure. Assuming that this local configuration is part of a well-defined simplicial manifold, this is no longer the case
after the move. After the  move, the vertices 1 and 3 are connected by two distinct spacelike edges, which violates the manifold conditions
we described earlier in Sec.\ \ref{sec:basis}. (The \emph{link} of the edge (03) after the move is not a circle topologically.) 

Assuming that the triangulation is a simplicial manifold before the update, the \code{flip} move does not lead to violations of the simplicial manifold conditions provided that:
\begin{enumerate}
    \item The vertices 0 and 2 both have spatial coordination number (i.e. the number of vertices connected to the vertex by spacelike links) of at least 4.
    \item There is no link connecting vertex 1 to vertex 3. (Note that such a link is always spacelike.)
\end{enumerate}
We enforce the first requirement by keeping track of the spatial coordination number of every vertex \code{v}, updating it each time a move changes the number of spatial links connected to \code{v}. If either of the vertices 0 or 2 have a spatial coordination number below 4 when the \code{flip} move is proposed, we reject the move. The second requirement is more computationally intensive to enforce, since we do not keep track of all links between vertices during the Monte Carlo simulation (we elaborate on this design choice in Sec.\ \ref{ssec:recon}). Prior to performing a \code{flip} move, we must therefore loop through all tetrahedra containing the vertex 0, rejecting the move if any of these tetrahedra are found to also contain the vertex 3.

\begin{figure}[t]
    \centering
    \includegraphics[width=0.7\textwidth]{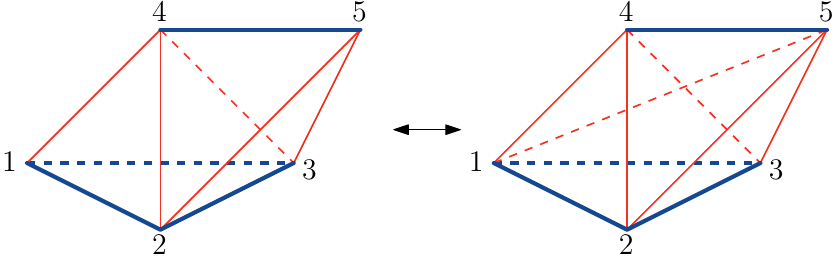}
    \caption{The \code{shift} and \code{ishift} moves for a (3,1)-simplex in 3D CDT.}
    \label{mc-fig:3d-cdt-shift-ishift}
\end{figure}

\subsubsection{The \code{shift} and \code{ishift} moves}
The input for the \code{shift} move is either a (3,1)- or a (1,3)-simplex. Both cases should be implemented to guarantee ergodicity, but since their 
description is analogous we restrict our attention here to the case of a (3,1)-simplex. We select this (3,1)-simplex, called \code{t}, from a given labelled triangulation $T$ with uniform probability $1/N_{31}(T)$. Subsequently we select with uniform probability $1/3$ one of the three neighbouring tetrahedra that share a timelike 
face with \code{t}, and denote it by \code{ta}. If \code{ta} is a (3,1)-simplex we abort the move. Conversely, if \code{ta} is a (2,2)-simplex, we propose 
a change to the geometry which replaces the timelike triangle shared by \code{t} and \code{ta} by its dual timelike edge.
The interior of the local configuration is rearranged accordingly, which involves the addition of a (2,2)-simplex, but its boundary is left unchanged.
In terms of the vertex labels of Fig.\ \ref{mc-fig:3d-cdt-shift-ishift}, we have $\code{t}=(1234)$ and $\code{ta}=(2345)$. During the move $T \rightarrow T'$, 
the interior triangle (234) is replaced by its dual timelike edge (15). At the same time,
the simplex pair of \code{t} and \code{ta} is replaced by a (3,1)-simplex $\code{tn0}=(1235)$ and two (2,2)-simplices, $\code{tn1} =(1245)$ and 
$\code{tn2} =(1345)$ as illustrated. No vertices are added during a \code{shift} move, and no vertex labels are rearranged. 

The combinatorial factor in the probabilities $P_l$ of eq.\ \eqref{mc-eq:prob-labeled} is the same for both $T$ and $T'$, but 
the addition of a (2,2)-simplex means that $N_3(T') = N_3(T) + 1$, which changes the action. 
The selection probability for this move is $g(T \to T') = \frac{1}{3 N_{31}(T)}$.

The inverse move, the (3,2)-move or \code{ishift} move, also takes as input a uniformly selected (3,1)-simplex \code{t}, now from the labelled triangulation $T'$. 
We then pick two of its neighbours, denoted \code{tn0} and \code{tn1}. The order in which they are picked is irrelevant, so this choice can be made in three distinct ways. The move is aborted unless \code{tn0} and \code{tn1} are neighbours and are both (2,2)-simplices. 
Otherwise, the timelike edge shared by \code{t}, \code{tn0} and \code{tn1} is replaced by its dual timelike triangle, and the three-simplices are
rearranged in a manner inverse to what was done during the \code{shift} move. The overall effect is to remove a (2,2)-simplex.
The selection probability for this move is again $g(T' \to T) = \frac{1}{3 N_{31}(T)}$. 
Taking into account the change in the action we find from eqs.\ \eqref{mc-eq:acceptance-ratios_1} and \eqref{mc-eq:acceptance-ratios_2} that 
\begin{align}
    A(T \to T') &= \textrm{min}\left(1, e^{-k_3} \right), \\
    A(T' \to T) &= \textrm{min}\left(1, e^{k_3} \right).
\end{align}
Note that the \code{shift} and \code{ishift} moves do not affect the geometry of the spatial slices, but only the interpolating 3D geometry 
between adjacent spatial slices.

\subsection{Fixing the volume}
\label{mc-sec:vol-fix}
We often want to collect measurements at a certain fixed discrete target volume $\tilde{N}$, especially in the context of a finite-size scaling analysis. However, there is no volume-preserving set of moves that is ergodic in the space of CDT geometries of fixed size, which means that we must allow the volume to fluctuate around $\tilde{N}$. It is difficult to tune the cosmological coupling parameter to its pseudocritical value associated with a certain target volume. Even then, the fluctuations around the target volume are large, and we only rarely encounter systems of the right size. A standard solution is to include a volume-fixing term $S_\textit{fix}$ in the action, which penalizes configurations that stray too far from the target volume. A convenient choice is a quadratic volume-fixing term of the form
\begin{equation}
    S_\textit{fix} = \epsilon \big( N-\tilde{N}\big)^2,
\label{volfix}
\end{equation}
where $N$ is the current system size and $\epsilon >0$ sets the strength of the fixing. When this term is taken into account in the computation of acceptance ratios and the cosmological coupling is tuned to its pseudocritical value, the volume tends to fluctuate around the target $\tilde{N}$ with a typical fluctuation size determined by $\epsilon$ (with larger $\epsilon$ leading to smaller fluctuations, and vice versa).

\section{Implementation}
\label{sec:impl}
In Sec.\ \ref{mc-sec:mcmrg} we explained the theoretical aspects of the Monte Carlo simulations of CDT, together with an explicit discussion of the
moves in two and three dimensions and the associated detailed-balance equations.  
In this section, we will discuss some of the challenges and technical issues that arise when putting these ideas into practice,
and the strategies we use to address them in the implementation of the simulation codebases \cite{brunekreef2021jorenb,brunekreef2022jorenb}. 
We do not aim to provide a full description of the C++ code itself, but rather focus on the generalities of our approach. We will use terminology from object-oriented programming (such as classes, objects, member functions, and parent/child inheritance) throughout this section, and we refer to \cite{lafore1997objectoriented} for a more detailed explanation of these concepts in the context of C++. 

In both codebases, the most important \emph{classes} from the user's point of view are \code{Universe} and \code{Simulation}. The class \code{Universe} represents the current state of the triangulation and stores properties of the geometry in a convenient manner. It also provides \emph{member functions} that carry out changes on the geometry. The \code{Simulation} class is responsible for all procedures related to the actual Monte Carlo simulation. It proposes moves and computes the detailed-balance conditions. If it decides a move should be accepted, it calls on the \code{Universe} to carry out the move at a given location. It also triggers the measurement of observables when the time is right. The user can define custom observables as \emph{children} (again in the sense of object-oriented programming) of the standard \emph{parent class} \code{Observable}. This class provides some of the basic functionality that one may require when implementing certain observables, like a breadth-first search algorithm for measuring distances, or constructing metric spheres on the (dual) lattice.

Two important classes operating mainly ``under the hood'' are the \code{Pool} and \code{Bag} classes. They are responsible for the memory management for all the simplices present in the triangulation and the user can generally consider them as black boxes. Since it is nevertheless useful to understand their capabilities and limitations, we 
will explain them in more detail below.

The remainder of this section covers several topics. In Sec.\ \ref{ssec:storage} we discuss how to store and access a triangulation and its constituents in the computer memory. The approach used is largely independent of the dimension of the triangulation, but dimension-specific considerations are required when implementing the Monte Carlo moves that generate a random walk in the space of triangulations. These are dealt with in Sec.\ \ref{ssec:recon} for both 
2D and 3D CDT. In Sec.\ \ref{ssec:simul} we provide further details on the various stages of the simulation process: 
the tuning of the coupling constants, the thermalization of the system, and the collection of measurements. Finally, in Sec.\ \ref{ssec:observables} we discuss the inclusion of observables, 
including a brief mention of the breadth-first search (BFS) algorithm used for measuring distances on the lattice. 

\subsection{Storage and access}
\label{ssec:storage}
Our goal is to store a $d$-dimensional triangulation $T$ in computer memory. The triangulation consists of a collection of $d$-simplices $\sigma^{(d)}\! \in\! T$, 
where each $\sigma^{(d)}$ has $d+1$ neighbouring $d$-simplices.\footnote{Note that we only  consider triangulations without boundary.} 
We can also organize the triangulation into collections of $k$-simplices, where $k = 0, 1, \cdots, d$. A $k$-simplex can be described by a list of
$k+1$ vertices (0-simplices). In a simplicial manifold, this description is unique in the sense that a given set of $(k+1)$ vertices cannot be associated 
with more than one $k$-simplex.

For a triangulation $T$ with $N_0$ vertices, a simple choice of labelling is to assign to each vertex $v \in T$ a distinct label from the set 
$\left\{0,1, \cdots, N_0-1\right\}$. 
Each $k$-simplex $\sigma^{(k)}\! \in\! T$ can then be denoted by an ordered $(k+1)$-tuple of its vertex labels.
As an example, consider the smallest manifold triangulation of a two-sphere, the surface of a tetrahedron, which contains four vertices with labels $\{0,1,2,3\}$. 
It also contains six 1-simplices or links (edges) and four 2-simplices or triangles, each of which has the other three triangles as its neighbours. 
The triangulation and the lists of its 0-, 1-, and 2-simplices are shown in Table \ref{mc-fig:tetrahedron}.
\begin{table}[ht!]
    \begin{minipage}{0.35\textwidth}
    \centering
    \includegraphics[width=0.55\textwidth]{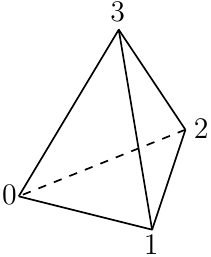}
    \end{minipage}
    \begin{minipage}{0.62\textwidth}
    \begin{tabular}{p{0.2\textwidth}|p{0.7\textwidth}}
    vertices & \{(0),(1),(2),(3)\} \\
    \hline
    links & \{(01), (02), (03), (12), (13), (23)\} \\
    \hline
    triangles & \{(012), (013), (023), (123)\}
    \end{tabular}
    \end{minipage}
    \caption{Minimal triangulation of a two-sphere, with lists of its $n$-simplices.}
    \label{mc-fig:tetrahedron}
\end{table}
The list of triangles is in principle sufficient to reconstruct the entire triangulation. For example, to find the neighbours of a given triangle \code{t}, we should search for all triangles that have two vertices in common with \code{t}. We can also make changes to the triangulation, by updating the list of triangles, and adding or removing vertices. For example, we can subdivide the triangle $(013)$ into three triangles by inserting a new vertex with label 4 in its centre. 
This corresponds to removing $(013)$ and adding three new triangles $(014), (034), (134)$ to the list of triangles.

While this strategy is sound in principle, it is inefficient in practice. Finding the neighbours of a triangle \code{t} requires a search of the entire list of triangles until we have found three objects that match the criterion of having two vertices with \code{t} in common. The average time duration of such a search scales linearly with the number $N_2$ of triangles in the system. In the language of complexity theory, we say that the time complexity of this operation is $O(N_2)$.
Recall that we want to construct large triangulations in order to approach the continuum limit and perform very large numbers of local updates on them. 
When performing a single local update, it is therefore imperative to avoid operations that scale with some (positive) power of the system size. 
This can be elevated to a general guiding principle for the simulation code:

\begin{mdframed}[align=center,userdefinedwidth=0.8\textwidth,skipabove=0.5cm]
\begin{center}
    \textbf{Operations required for performing Monte Carlo moves on the geometry should run in $O(1)$ time.}
\end{center}
\end{mdframed}
\vspace{0.4cm}
The big-O notation should again be understood as is customary in complexity theory: an operation on a $d$-dimensional CDT geometry is said to run in $O(1)$ time --- also referred to as \emph{constant time} --- if its execution time does not depend on the total number $N_d$ of $d$-simplices in the system. The requirement of constant execution time per basic operation precludes the use of searches of the full lists of simplices. 

One may expect that a straightforward object-oriented approach can solve this issue. For example, in the two-dimensional case we could formulate three classes of simplex objects, \code{Vertex}, \code{Link} and \code{Triangle}. A \code{Triangle} object could then store lists with references to the three \code{Vertex} objects and the three \code{Link} objects that are contained in it, and a list of references to the three \code{Triangle} objects adjacent to it. This is a step in the right direction, but not yet a solution to all our problems. The reason is that the analogous lists of references stored by \code{Vertex} and \code{Link} objects are not all of constant size. A single vertex can be a member of an arbitrary number of triangles, which implies that the list of triangles that contain a given vertex is unbounded in length.\footnote{Strictly speaking it is bounded in terms of the total system size, but the difference is immaterial for our argument.} An example of a situation where this is problematic is the execution of a \code{delete} move in 2D CDT. Two triangles are removed from the system, and three vertices (labelled $1,2,3$ in Fig.\ \ref{mc-fig:2d-cdt-moves}, top) should be updated to reflect this change. This would again require a search of the list of references to \code{Triangle} objects stored in the \code{Vertex} object, slowing down the procedure.\footnote{An additional motivation for avoiding the use of variable-length lists in the simplex objects is that they can incur an overhead in allocation time compared to fixed-length lists. However, advances in computer architecture and compiler design have brought down this overhead significantly, making this perhaps only a minor point of concern.}

The upshot is that we should find a strategy for storing a certain minimal amount of information required to perform all desired operations on lists of simplex objects in $O(1)$ time. It is convenient to maintain separate lists of simplices according to the simplex dimensionality $k$.  
In what follows we will work with a fixed, but arbitrary $k\leq d$. 

Four basic operations are required for Monte Carlo simulations of CDT:
\begin{enumerate}
    \item \textbf{Object creation:} add a $k$-simplex to a given triangulation and assign to it a unique label, thereby increasing the total list size $N_k$ by one.
    \item \textbf{Object deletion:} given the label of a $k$-simplex, remove the simplex from the triangulation, thereby decreasing the total list size $N_k$ by one.
    \item \textbf{Object lookup:} given the label of a $k$-simplex, find the simplex in memory and access its properties.
    \item \textbf{Random object selection:} pick a random $k$-simplex (potentially satisfying certain conditions) in the triangulation with uniform probability.
\end{enumerate}
Combining these four operations is what makes the implementation nontrivial, as we will see in due course. It turns out that in dimension $d=2,3$ the operations 1-4 are required by moves only for $k=0$ and $k=d$, that is, for the vertices and the simplices of top-dimension.
We will now proceed to explain the \code{Pool} structure, which is used to implement the operations 1--3 in our codebase, and then treat the \code{Bag} structure, which implements operation 4.

\subsubsection{The \code{Pool} structure}
A straightforward approach for managing simplices would be to store them in a contiguous array. The deletion of an arbitrary object from such an array leaves behind a hole. One can fill it by moving the last element \code{obj} of the array to the newly vacant position, but this introduces a new problem: all other objects that refer to \code{obj} should be updated to reflect the change of its position in the array, and this updating operation is potentially of $O(N_k)$ complexity. An alternative solution is to include hole formation in our strategy, and work with an array of fixed length. This fixed length $C$ imposes an upper bound on the number of simplices that can be created, but memory capacity is typically not a bottleneck in CDT simulations, where a computing node with 4GB of memory can easily store on the order of millions of simplices. A larger concern with systems of this size often is their long thermalization time. Our strategy will be to work with vastly expanded arrays of length $C\gg N_k$, but where only 
$N_k$ array elements are ``used'', in a sense we will describe below.

The \code{Pool} data structure implements the idea of an array with ``holes''. A \code{Pool} is instantiated with a user-defined capacity $C$, allocating $C$ objects in memory, with objects stored in the array \code{Pool::elements}. The elements persist in memory throughout the simulation, and are only deallocated at the end of the simulation run. The choice of a value for $C$ is a trade-off between the available memory and the expected maximal
simplex numbers during all stages of the simulation, including thermalization. For example, in the 2D simulations we use $C=10^7$ for vertices and $C=2\times 10^7$ for triangles.\footnote{The reason for this factor 2 difference is that there are twice as many triangles as vertices in an arbitrary two-dimensional triangulation of topology $S^1 \times S^1$.}

Objects in the \code{Pool} are identified uniquely by their index \code{i}, where $\code{i}\!\in\! [0,C-1]$, such that an object with index \code{i} is located at
the $(\code{i}+1)$th position of the array \code{Pool::elements}. Two additional pieces of information are stored with every element in the list, a binary 
variable $x$ and a \code{Pool} index variable $p\!\in\! [0,C-1]$. If $x[\code{i}]\! =\! 1$ at a given time, the object stored at index \code{i} is ``used'' and corresponds to 
a $k$-simplex with label \code{i} contained in the triangulation at that moment, 
while $x[\code{i}]\! =\!0$ means it is ``free''. 
The functionality of the variable $p[\code{i}]$ will be explained in due course. 
In addition to these quantities, the user can define other custom member variables for objects in the \code{Pool}. 
Lastly, we need one other variable \code{first} for the entire system, which takes values in $[0,C-1]$ and keeps
track of the index of the (free) object that is next in line for being converted to used status.

To create an object from the point of view of the simulation, we call the function \code{Pool::create()}. 
This takes as input the index given by \code{first}, some value \code{j}, say, and switches the status of the object with index \code{j} to used
by setting $x[\code{j}]\! =\! 1$. The variable \code{first} is then updated to the new value $\code{first}\! =\! p[j]$, pointing it to the next free object, which is 
next in line for being converted to ``used'' if \code{Pool::create()} is called again. This defines and explains the functionality of the variable $p$ introduced above.

Initially, all objects are marked free, $x[\code{i}]\! =\!0$, as well as $p[\code{i}]\! =\!\code{i}\! +\! 1$ for all indices \code{i}, and $\code{first}\! =\! 0$, linking all objects in the array \code{Pool::elements} into a list. 
To illustrate the general process we just described, let us create a sequence of objects from this starting point.
Since $\code{first}\! =\! 0$, it amounts to first picking the free object with index 0, setting $x[0]\! =\! 1$, and updating 
$\code{first}=p[0]\equiv 1$. The first object in \code{Pool::elements}, with label $0$, is then marked ``used''. Since $\code{first}= 1$, the (free) object next in line
has index 1. We set $x[1]=1$ and update $\code{first}=p[1]\equiv 2$, thereby creating a new object with label 1, and so forth. 

Similarly, we can destroy an object with label \code{i} from the point of view of the simulation by calling \code{Pool::destroy(i)}. 
This marks the object as free by setting $x[\code{i}]\! =\! 0$. The index \code{i} is moved to the ``top'' of the sequence of indices of free objects
waiting to become used again, by 
setting $p[\code{i}]\! =\!\code{first}$ and then $\code{first}\! = \!\code{i}$.
Before destroying an object, the \code{Pool} checks whether the object is indeed free. If not, an error is thrown, since an attempt by the simulation
to destroy a free object signals a bug in the code.

Accessing an arbitrary object is also straightforward, where the function \code{Pool::at(i)} returns the object with label \code{i}. 
To summarize, the \code{Pool} structure provides us with object creation, deletion, and lookup operations with $O(1)$ complexity. 
However, the fact that the list \code{Pool::elements} generically contains holes, in the sense of free objects,
makes the random selection of an object in $O(1)$ time impossible. 
It requires an extra data structure, which tracks the position of all used objects in the \code{Pool}. This is the \code{Bag} class, which we will discuss next.

\subsubsection{The \code{Bag} structure}
It is straightforward to pick a random object from a contiguous list of length $N$: generate a random integer $i$ in the range $[0, N\! -\! 1]$ and retrieve the object at index $i$ in the list. However, this does not work for the list \code{Pool::elements}, since an index may point to a free object. For the same reason, we cannot iterate over all used objects (or a subset of them) present in the \code{Pool}. The \code{Bag} class is a structure that can be tied to a \code{Pool} to provide this missing functionality.

The \code{Bag} contains two arrays of fixed length $C$, called \code{Bag::elements} and \code{Bag::indices},
where $C$ is equal to the associated \code{Pool} capacity. At a given moment, either list contains $n$ ``active'' entries, associated with objects that are used, 
and $C\! -\! n$ inactive ones, where all inactive lists entries are set to the \code{EMPTY} marker $-1$. We keep track of the number of active entries by
the variable $\code{Bag::size}\equiv n$.

Note that the \code{Bag} does not contain the objects themselves, but rather is a bookkeeping device for their labels.
The first $n$ elements of the list \code{Bag::elements} contain these labels in some order. The important aspect is that this part of the list is 
\emph{contiguous}, in the sense that it does not have any ``holes'' (inactive elements). 
The list \code{Bag::indices} has as its (active) entries the indices (list positions in \code{Bag::elements}) of those labels: the entry 
$\code{j}=\code{Bag::indices}[\code{i}]$ is the index in the list \code{Bag::elements} which contains the label \code{i}. 
The two lists are dual to each other in the sense that 
\begin{center}$\code{Bag::elements[Bag::indices[i]] = i}$\,\, \end{center}
\begin{center}$\code{Bag::indices[Bag::elements[j]] = j}$. \end{center}
To reiterate, the array \code{Bag::elements} is contiguous with indices $0, \cdots, n-1$ occupied, while \code{Bag::indices} is indexed by labels of $k$-simplices which do not change throughout the whole lifetime of the simplex.

Such a structure allows us to add and remove objects to/from a \code{Bag} in $O(1)$ time. 
When an object is added using \code{Bag::add(i)}, its label \code{i} is entered at \code{Bag::elements[Bag::size]}, 
the element \code{Bag::indices[i]} is set to \code{Bag::size}, and \code{Bag::size} is increased by 1.
Removal of an object with label \code{i} using \code{Bag::remove(i)} is equally straightforward. We look up the index stored at \code{Bag::indices[i]} pointing to the location \code{j} of \code{Bag::elements}, and replace the label stored at location \code{j} 
by the last non-empty element of \code{Bag::elements} (the one located at position $\code{Bag::size}-1$), to avoid creating a ``hole'' at index \code{j}.
We then update the index of this last element in \code{Bag::indices}, and set both $\code{Bag::elements}[\code{Bag::size} - 1]$ and \code{Bag::indices[i]} to $-1$.
Finally, \code{Bag::size} is decreased by 1.

We can now also select a random (active) element from a \code{Bag}, by gene\-rating a random integer \code{j} in the range $[0, \texttt{\code{Bag::size}}-1]$ and retrieving the label of the object located at \code{Bag::elements[j]}. This is implemented by the function \code{Bag::pick()}. Similarly, we can iterate over all object labels in the \code{Bag} by restricting iteration over \code{Bag::elements} from index 0 to $\code{Bag::size}-1$. 
Once we have obtained an object label from a \code{Bag}, we can retrieve the properties of the corresponding object by accessing it in the associated \code{Pool}.

We can have more than one \code{Bag} for a given object type (like \code{Vertex} or \code{Triangle}) and the corresponding \code{Pool}.
This is convenient whenever we work not only with the complete set of $k$-simplices, but also with a subset that satisfies additional conditions, as happens
when using the bookkeeping method.
As an example, take the 2D \code{add/delete} moves of Sec.\ \ref{sub:adddelete} in the bookkeeping version. 
For the \code{add} move, which needs an arbitrary \code{Triangle} as input, we can define a \code{Bag trianglesAll} from which we draw random triangles with \code{trianglesAll.pick()}. The \code{delete} move requires a \code{Vertex} object of order four, for which we can define a \code{Bag verticesFour} containing all vertices of this particular type. To maintain consistency, 
we should take care to keep this \code{Bag} updated throughout the simulation, while adding and removing vertices. 
Finally, we can keep a \code{Bag trianglesFlip} containing all \code{Triangle} objects that can be used as an input for the \code{flip} move described in Sec.\ \ref{sub:flip}.

Since every \code{Bag} allocates two arrays of integers, both of fixed length equal to the total capacity $C$ of the corresponding \code{Pool}, it requires a large amount of memory and we should keep the number of distinct \code{Bag}s to a minimum. 
This is also why we cannot simply add a \code{Bag} to every \code{Vertex} object, to keep track of its variable number of neighbours. 

\subsection{Geometry reconstruction}
\label{ssec:recon}
We have shown above how to efficiently store and access triangulation data in the computer memory. One of our findings was that this precludes the use of 
lists of variable length, like those tracking the neighbours of a given vertex. However, since the measurement of some observables requires this information, we should be able to reconstruct it whenever needed. To further illustrate the point, we saw that the \code{delete} move in 2D takes a \code{Vertex} of order four as input and subsequently deletes two \code{Triangle} objects. This suggests that \code{Vertex} should contain \emph{some} information on its neighbouring simplices. We will show next how to keep track of a minimal amount of extra information that suffices for our purposes. There is no unique way of accomplishing this. Here we will restrict ourselves to the approach taken in our simulation code, treating the cases of two and three dimensions separately.

\subsubsection{Reconstruction in 2D CDT}
\begin{figure}[t]
    \centering
    \includegraphics[width=0.8\textwidth]{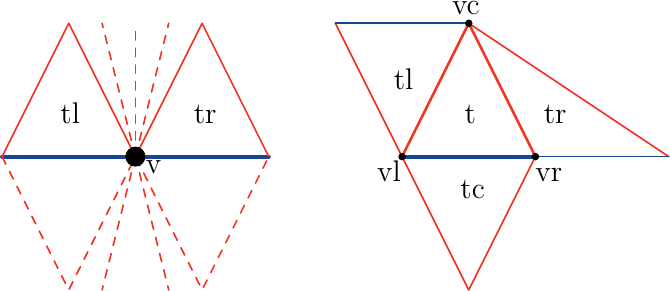}
    \caption{Schematic representation of the local connectivity data stored in \code{Vertex} (left) and \code{Triangle} (right) objects in 2D CDT.}
    \label{mc-fig:neighborhoods-2d}
\end{figure}

There is a simple solution in 2D, which allows us to reconstruct the complete connectivity information of a triangulation, while preserving $O(1)$ execution time for individual moves. By this we mean all data about neighbourhood relations, for example, all neighbouring vertices of a given vertex, or all triangles that share a given vertex. As mentioned earlier, we keep track of two fundamental simplex types during updates of the geometry in 2D, the \code{Vertex} and the \code{Triangle}. 
Alongside each such object, we store the following information (cf.\ Fig.\ \ref{mc-fig:neighborhoods-2d}):
\begin{itemize}
    \item \code{Vertex}
    \begin{itemize}
        \item \code{Triangle tl, tr;} the two (2,1)-simplices containing the vertex \code{v}, to the left and right of \code{v}.
    \end{itemize}
    \begin{samepage}
    \item \code{Triangle}
    \begin{itemize}
        \item \code{Vertex vl, vr, vc;} the three vertices comprising the triangle \code{t}. The vertices \code{vl} and \code{vr} lie at the left and right end of the base of \code{t} (in the same spatial slice), and \code{vc} at its apex.
        \item \code{Triangle tl, tr, tc;} the three triangles sharing an edge with the triangle \code{t}. The triangle \code{tl} is on the left, \code{tr} is on the right, and \code{tc} is its vertical neighbour.
    \end{itemize}
    \end{samepage}
\end{itemize}
Using this information alone, all moves described in Sec.\ \ref{mc-fig:2d-cdt-moves} can be implemented with $O(1)$ execution time, and all other neighbourhood relations can be reconstructed. For illustration, let us show how to obtain a list of all vertex neighbours of a given \code{Vertex v}: 
start from the \code{Triangle v.tl}, and add all of its vertices (except \code{v}) to the collection. Then step to the triangle on its right using \code{v.tl.tr}, and add its one vertex that is not shared with the previous triangle. Continue moving to the next triangle neighbour on the right and collecting vertices, until arriving at \code{v.tr}. Then step down using \code{v.tr.tc}, and complete the circle by stepping to the left, each time adding a new vertex, until getting to \code{v.tl.tc}. Similarly, one can reconstruct a list of all \code{Link} objects in the triangulation by iterating through the list of triangles. This shows that it would have been redundant  
to keep this list up-to-date at every step of the Monte Carlo simulations. 
It is more efficient to perform the full reconstruction at the time we want to take a measurement. 

\subsubsection{Reconstruction in 3D CDT}
\begin{figure}[t]
    \centering
    \includegraphics[width=0.8\textwidth]{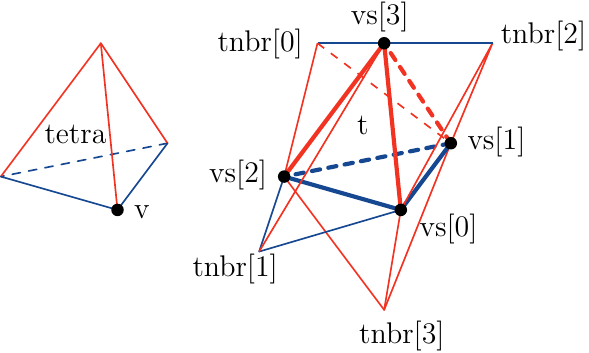}
    \caption{Schematic representation of the local connectivity data stored in \code{Vertex} (left) and \code{Tetra} (right) objects in 3D CDT.}
    \label{fig:neighborhoods-3d}
\end{figure}

Similar considerations apply to the 3D version of the model. The fundamental objects that are tracked during updates of the geometry are the \code{Vertex} and the \code{Tetra} (short for tetrahedron). They store the following geometric data, as presented graphically in Fig.\ \ref{fig:neighborhoods-3d}:
\begin{itemize}
    \item \code{Vertex}
    \begin{itemize}
        \item \code{Tetra tetra;} an arbitrary (3,1)-simplex containing the vertex \code{v} in its base (i.e.\ not as its apex).
        \item \code{int cnum;} the coordination number of the vertex \code{v}, i.e.\ the number of tetrahedra containing it.
    \end{itemize}
    \item \code{Tetra}
    \begin{itemize}
        \item \code{Vertex[4] vs;} an array of the four vertices that comprise the tetrahedron \code{t}.
        \item \code{Tetra[4] tnbr;} an array of the four tetrahedra sharing a face (triangle) with the tetrahedron \code{t}.
    \end{itemize}
\end{itemize}
We use two ordering conventions for the lists associated with the \code{Tetra} object, which allow us to retrieve specific (sub-)simplices based on their relation to the given tetrahedron \code{t}. First, \code{t.tnbr[i]} (with $\texttt{\code{i}} \in \{0,1,2,3\}$) refers to the neighbouring tetrahedron of \code{t} which lies \emph{opposite} to the vertex \code{t.vs[i]}. Equivalently, \code{t.tnbr[i]} is the tetrahedron that shares with \code{t} the triangle spanned by the three vertices in \code{t.vs} which are \emph{not} equal to \code{t.vs[i]}.

Second, we (partially) order the vertices in the list \code{Tetra::vs} according to increasing $\tau$, where $\tau$ is the discrete time label of the spatial slice that contains a given vertex. This implies that the ordering depends on the tetrahedral type. For tetrahedra that lie between adjacent spatial slices at times $\tau$ and $\tau+1$,  
the time labels of the vertices in \code{t.vs} of a (3,1)-simplex \code{t} are $[\tau,\tau,\tau,\tau\! +\! 1]$, those for a (1,3)-simplex are $[\tau,\tau\! +\!1,\tau\! +\! 1,\tau\!+\! 1]$, and those for a (2,2)-simplex
are $[\tau,\tau,\tau\! +\!1,\tau\! +\!1]$. There is no particular ordering for vertices with identical time labels.

All moves described in Sec.\ \ref{mc-sec:3d-cdt-moves} can be performed in $O(1)$ time, given the information listed above. This is immediately clear for the \code{add}, \code{shift}, \code{ishift}, and \code{flip} moves. To achieve $O(1)$ complexity also for the 
\code{delete} move, we keep track of the vertex coordination numbers by storing them in \code{Vertex::cnum} throughout the simulation.\footnote{Updating this information continuously is not free, and incurs a small overhead in execution time. However, large vertex coordination numbers are relatively common for sufficiently small couplings $k_0$, and computing the coordination number on the fly using a neighbourhood reconstruction for such vertices is expensive. We therefore estimate that the simple operation of updating several integer variables in every move is more efficient.} It is then straightforward to check whether a randomly selected \code{Vertex v} is a candidate for the \code{delete} move, since it qualifies if and only if it is of order six. The tetrahedra involved in the move can be found by a neighbourhood reconstruction similar to the one described for the 2D model, where \code{Vertex::tetra} is now taken as a starting point. The same algorithm can be used to reconstruct all information about the connectivity of the triangulation whenever we want to perform measurements.

\subsection{Simulation stages}
\label{ssec:simul}
A single simulation run can generally be divided into three stages:
\begin{enumerate}
    \item \textbf{Tuning.} Adjusting the bare cosmological coupling parameter to its pseudocritical value, which will typically depend on the values chosen for the remaining coupling constants of the model.
    \item \textbf{Thermalization.} Evolving the system from the initial input geometry to an independent, thermalized configuration.
    \item \textbf{Measurement.} Collecting measurement data of geometric observables on snapshots of configurations occurring in the random walk.
\end{enumerate}
Separating these stages is a matter of convention, and there are no strict criteria that signal when the next stage of the simulation should be entered. The tuning stage is not relevant in pure 2D CDT, where the exact value $\lambda^{\textrm{crit}}$ of the critical cosmological coupling for the simplicial-manifold ensemble is known. In models where we require a tuning of the cosmological coupling, this tuning phase partially thermalizes the system. In some set-ups the tuning phase is never left, and the coupling is continuously tuned even while measurements are being taken. Finally, there is no binary switch that tells us exactly when the system has thermalized. Deciding when to start the measurement phase is a mix of art and science. We will now discuss briefly the global features of the three simulation stages. 

\subsubsection{Tuning}
For a general model of random geometry, it is unknown how to analytically compute the critical value of the bare cosmological coupling parameter as a function of the remaining couplings. An exception is 2D CDT, where $\lambda^{\textrm{crit}} \! =\! \ln 2$ exactly. Coupling the pure-gravity model to matter will change this critical exponent, but few numerical and analytical results are currently available \cite{ambjorn1999new,ambjorn2009shaken,durhuus2021critical}.  

In other cases, like CDT in three and four dimensions, we have to estimate the (pseudo-)critical value of the coupling by manually adjusting it in small steps until we find a range where the system approaches the limit of infinite volume. This so-called \emph{tuning} of the coupling proceeds roughly speaking as follows. 
In any dimension $d$, denote the bare cosmological coupling by $\lambda$ and pick a target volume $\tilde{N}$. Guess an arbitrary initial value for $\lambda_0$, typically of order 1, and initialize a Monte Carlo simulation of the system, using $\lambda_0$ as the cosmological parameter when computing acceptance ratios for the moves. Let the simulation run for a sufficiently large number of moves, keeping track of the total system volume at set intervals. Next, compute the average $\langle N \rangle$ of all system volumes recorded during this period, and adjust $\lambda$ upwards if $\langle N \rangle\! >\! \tilde{N}$, and downwards if $\langle N \rangle\! < \! \tilde{N}$. Continue the simulation with the adjusted value $\lambda_1$ as the cosmological coupling. Iterating this procedure produces a sequence of values $\lambda_i$, which approach the pseudocritical value $\lambda^{\textrm{crit}}$ associated with the target volume $\tilde{N}$ and the remaining coupling parameters. We stop the tuning procedure when the average $\langle N \rangle$ is within a sufficiently small range of the target volume $\tilde{N}$.

An efficient method for approaching the pseudo-critical coupling is to make the tuning process \emph{adaptive}, in the sense that we choose larger adjustments to $\lambda$ if the absolute difference $\big| \tilde{N}\! -\!\langle N \rangle \big|$ is large, and smaller adjustments as the difference decreases. Finding appropriate step sizes for such an adaptive procedure must be done on a trial-and-error basis, since we do not know how to estimate them analytically. A tuning procedure for 3D CDT is implemented in our code by the function \code{Simulation::tune()}.

While tuning the coupling, it is useful to include a volume-fixing term as discussed in Sec.\ \ref{mc-sec:vol-fix}. It reduces the magnitude of fluctuations around the target volume, and thereby prevents the system from shrinking to zero size or exploding to infinite size whenever the initial guess $\lambda_0$ is too far from the true pseudocritical value.

\subsubsection{Thermalization}
Once we have found a suitable estimate of the pseudocritical value for the cosmological constant, the system volume fluctuates around the target volume $\tilde{N}$ with a typical fluctuation size that depends on the prefactor $\epsilon$ in the volume-fixing term (\ref{volfix}). We can now start the thermalization phase, which amounts to letting the simulation run sufficiently long for the random walk to bring us sufficiently far away from our starting geometry. As pointed out earlier, there are no strict definitions of ``sufficiently long'' and ``sufficiently far away'' that are applicable in a general context.

We group the moves during the thermalization and measurement phases in bunches called \emph{sweeps}, where one sweep corresponds to a fixed number of attempted moves. Typical sweep sizes are on the order of 10 to 100 times the target volume $\tilde{N}$. 2D CDT without matter does not require long thermali\-zation times, and it is often sufficient to let the thermalization run for about 100 sweeps. Higher-dimensional models tend to thermalize much more slowly, with a thermalization speed that depends on the location in phase space. We therefore prefer to set the length of the thermalization phase (in terms of the number of sweeps) based on an analysis of the measurement data, for example, by checking whether the value of some simple order parameter has reached a stable average with fluctuations around it.\footnote{Note that this method is not appropriate when measuring near a first-order phase transition, where the order parameters can exhibit discontinuous jumps between two distinct values.} When computing ensemble averages of measurements, we simply remove the initial region of the measurement phase.

\subsubsection{Measurement}
When thermalization has been completed, we enter the measurement stage, where data are collected to compute estimators of ensemble averages $\langle \mo \rangle$ of observables $\mo$. At every step in the measurement phase, we first let the system evolve for a certain amount of time, and then collect one data point for each observable included in the simulation. These data points are written to external files, allowing the user to analyze them before the full simulation run has been completed.

Just as for the thermalization, the appropriate sweep size $s$ and number $k$ of sweeps between measurements depend on the situation. The relevant quantity here is the \emph{autocorrelation time} for a given observable under investigation. It is expressed in terms of the number of attempted Monte Carlo moves and measures how long the system should thermalize between subsequent measurements of the observable, in order that these measurements can be considered statistically independent. We refer to \cite{newman1999monte} for a more detailed discussion of the computation of autocorrelation times in Monte Carlo simulations.

A slightly adapted strategy is needed when performing measurements at a fixed volume, which is typically the case in a finite-size scaling analysis. After $k$ sweeps, the system will then usually not have the exact target volume $\tilde{N}$. A simple solution is to complete $k$ sweeps first, continue the random walk until the system hits the size $\tilde{N}$, and only then perform a measurement. The number $k \cdot s$ is therefore the \emph{minimum} number of attempted moves between subsequent measurements, while the actual number of attempted moves is generally larger.

Before embarking on the actual measurements, we must reconstruct all the connectivity data of the triangulation, as explained in Sec.\ \ref{ssec:recon}. 
As an example, some observables perform a breadth-first search algorithm on the vertices of a triangulation to construct metric spheres of a given radius, and this algorithm requires that we know what \code{Vertex} objects are connected to any given \code{Vertex} by a link in the triangulation. The reconstruction of this information is triggered in our code by calling the function \code{Simulation::prepare()}. After the reconstruction has been completed, all observables that were registered using \code{Simulation::addObservable()} will be measured in the current state of the geometry. 

\subsection{Observables}
\label{ssec:observables}
We can measure specific geometric properties of a triangulation by defining child classes of the generic parent \code{Observable} class. An instance of such an \code{Observable} can be registered by calling \code{Simulation::addObservable()}, which ensures that the observable is computed on the current state of the system at the end of each measurement step, and its output is written to a file.

Reconstruction of the geometry takes place before any measurements are taken, and its results are stored in indexed lists. For example, we can retrieve the collection of vertices that are direct neighbours of a given vertex with label \code{i} by calling \code{Universe::vertexNeighbors.at(i)}. We emphasize again that this information is not directly available during the Monte Carlo updating, which should be kept in mind when accessing such lists at intermediate stages.

The result of a measurement is encoded as a text string, and stored in the variable \code{Observable::output}. The content of this variable is automatically appended to an external data file after the measurement has been completed.

We can define distances between two vertices in the triangulation as the smallest number of links connecting these vertices. An analogous definition exists for distances between $d$-simplices, where the links are replaced by dual links connecting these $d$-simplices. We call these the \emph{link distance} and \emph{dual link distance}, respectively. Measuring such distances on the lattice can be achieved by a breadth-first search (BFS) algorithm. For the case of the link distance, such an algorithm starts from a given vertex \code{i} in the triangulation, and enumerates all its neighbours (i.e. vertices connected to it by a single link) recursively until it arrives at the target vertex \code{f}. The algorithm for dual link distances works analogously. 

Since distance-finding is often necessary when investigating CDT geometries, the \code{Observable} member functions \code{distance(i, f)} and \linebreak \code{distanceDual(i, f)} implement this procedure on both the triangulation and its dual, where \code{i} and \code{f} represent the initial and target simplex respectively. Similarly, metric spheres of radius \code{r} centred at a simplex \code{c} can be constructed (also using a BFS algorithm) by calling the \code{Observable} member functions \code{sphere(c, r)} and
\code{sphereDual(c, r)}.
The implementation of the BFS was chosen to be heavy on memory usage, with the advantage of improved speed. The reasons are similar to those given when implementing the \code{Bag} structure.

Several observables are provided in the open-source code, which the user can take as guiding examples when implementing their own custom observables. Both the 2D and 3D codebases include the volume profile, the shell volumes used for determining the Hausdorff dimension, and the average sphere distance used in the context of the quantum Ricci curvature.

\section{Summary and outlook}
\label{sec:summ}

After motivating the use of lattice methods in quantum gravity and discussing some basic aspects of how the nonperturbative path integral
is formulated in terms of Causal Dynamical Triangulations, we presented above a concrete and detailed account of the implementation of Monte Carlo methods in CDT quantum gravity in two and three dimensions. 
The dynamical character of the lattice leads to subtleties in the implementation of the Monte Carlo moves and detailed balance, 
and to challenges for data storage and retrieval, which are not present on fixed lattices. 
We showed how to deal with these issues in many explicit examples, and discussed which aspects of the implementation have to be weighed against each other 
when optimizing the efficiency of the simulation. Our treatment is a hands-on guide for practitioners and, in conjunction with the corresponding open-source codes  \cite{brunekreef2021jorenb,brunekreef2022jorenb}, provides an entry point for newcomers to Monte Carlo simulations in quantum gravity.  

Computer simulations based on CDT have already proven to be an invaluable asset for extracting the quantitative behaviour of 
diffeomorphism-invariant observables in a Planckian regime \cite{ambjorn2012nonperturbative,loll2019quantum,ambjorn2021cdt,ambjorn2023lattice,loll2023quantum}. 
In view of the scarcity of nonperturbative tools in quantum gravity, the role of such numerical methods is bound to increase further. 
Technological developments in computer hardware will expand the scope of CDT research and allow for the simulation of larger systems, 
improving our capability to study the continuum physics.

In addition, there are interesting avenues to pursue on the software side of CDT Monte Carlo simulations. For example, as already mentioned in Sec.\ \ref{mc-sec:mcmrg}, further algorithmic improvements may be uncovered by a deeper exploration of the bookkeeping method. This could be particularly 
useful in regions of phase space that are currently difficult to study numerically, like those near higher-order phase transition lines.

Another tantalizing possibility are recent advances in the field of artificial intelligence. Increased usage and availability of graphical processing units (GPUs) has boosted research in machine learning, raising the question whether CDT research could similarly benefit from these developments. 
At the present stage, it remains difficult to see how the Markov chain Monte Carlo methods used in CDT simulations can be adapted 
to make efficient use of GPU architectures. A different and perhaps simpler option may be to employ GPUs for the measurement of computationally expensive observables, like the average sphere distance. 

\section*{Acknowledgments}
This work was partially supported by the research program “Quantum gravity and
the search for quantum spacetime” of the Foundation for Fundamental Research
on Matter (FOM, now defunct), financially supported by the Netherlands Organisation for Scientific Research (NWO). The work was also partially supported by the National Science Centre, Poland, under research project no. 2019/33/B/ST2/00589, and by the Excellence Initiative - Research University Program at the Jagiellonian University in Krak\'ow.

We thank T. Budd and J. Ambj\o rn for valuable discussions on CDT simulations. Furthermore, we thank D. N\'emeth for his contributions to the 3D CDT codebase. 

\vspace{1cm}

\printbibliography

\end{document}